\documentclass[12pt]{article}
\usepackage{epsfig}
\usepackage{amsfonts}
\usepackage{amsmath}
\usepackage{amssymb}

\newcommand{\simge}{\hspace*{0.2em}\raisebox{0.5ex}{$>$}
     \hspace{-0.8em}\raisebox{-0.3em}{$\sim$}\hspace*{0.2em}}
\newcommand{\simle}{\hspace*{0.2em}\raisebox{0.5ex}{$<$}
     \hspace{-0.8em}\raisebox{-0.3em}{$\sim$}\hspace*{0.2em}}

\newcommand{\dslash}[1]{#1 \llap{/\kern-0.5pt}}
\newcommand{\Dslash}[1]{#1 \llap{/\kern+1.2pt}}
\newcommand{\DDslash}[1]{#1 \llap{/\kern+2.3pt}}
\newcommand{\dslashh}[1]{#1 \llap{/\kern+1pt}}

\newcommand{\boldtau}{\mbox{\boldmath $\tau$}}

\def\bdm{\begin{displaymath}}
\def\edm{\end{displaymath}}

\begin{document}

\begin{titlepage}

\vspace*{1.5cm}

\begin{center}
{\Large\bf Few-Nucleon Systems in a Quirky World:}\\
\vspace{.5cm}
{\Large\bf Lattice Nuclei in Effective Field Theory}

\vspace{2.0cm}

{\large \bf U. van Kolck}

\vspace{0.5cm}
{
{\it Institut de Physique Nucl\'eaire, Universit\'e Paris-Sud, CNRS/IN2P3
\\
91406 Orsay, France}
\\
and
\\
{\it Department of Physics, University of Arizona,
\\
Tucson, AZ 85721, USA}
}

\vspace{1cm}

\today

\end{center}

\vspace{1.5cm}

\begin{abstract}
I describe how nuclear structure can be predicted from lattice QCD through 
low-energy effective field theories, using as an example a world
simulation with relatively heavy up and down quarks. 
\end{abstract}

\vfill
\end{titlepage}

\section{Introduction}
\label{intro}
Weakly bound systems are fascinating for
the surprising quantum features they display regardless of the details
of their short-distance structure.
They beg for a description with effective field theory (EFT),
because that is the general framework to turn
separated physics scales into a controlled expansion based
on symmetries, rather than details of the dynamics.
I have been smitten with EFT since shortly after leaving 
S\~ao Paulo for my Ph.D. in Austin.
Only much later did I learn that a master of weakly bound systems
(halo nuclei \cite{Canto}, neutron-rich nuclei \cite{Bertulani:1993qt}, 
atoms near Feshbach resonances \cite{Timmermans},...)
lived in my hometown. This contribution is dedicated
to this master, 
Professor Mahir Saleh Hussein, on the occasion of his 70th birthday.

I will argue here, based on the work of Refs. 
\cite{latticenuclei,latticenuclei2,latticenuclei3},
that recent lattice QCD (LQCD) data 
\cite{Yamazaki12,NPLQCD13a,NPLQCD13b,Yamazaki15} 
suggest that light nuclei are weakly bound even in a world with
relatively large quark masses.
It seems that these data can be described with an EFT,
Pionless or Contact EFT  \cite{piless2,piless3,piless4}, 
where all degrees of freedom except nucleons
are implicit \cite{paulo}. 
This is, {\it mutatis mutandis}, the same theory that one can use
to describe other systems and processes characterized by relatively 
large sizes:
halo nuclei such as $^6$He \cite{6He}, 
shallow molecules such as He$_4$ trimers \cite{3atoms},
atom recombination near Feshbach resonances \cite{3fesh},
{\it etc.}
It has been known for some time to apply to light nuclei
at the physical value of the quark masses \cite{2N,3N,3Nscatt,4N,4Nscatt,6N}.
If more or less familiar nuclear structure can be predicted
for larger quark masses, perhaps nuclear physics 
is less accidental than we are used to think.

This is not just an academic exercise. 
As quark masses decrease in LQCD, pions become lighter and
another, less universal but more predictive EFT, 
Pionful or Chiral EFT \cite{paulo,nonEFT}, 
can be used to connect results at 
different quark masses \cite{BBSvK}. 
We will at that point be able to predict even more deeply bound
nuclei from LQCD, following the same steps as in Refs. 
\cite{latticenuclei,latticenuclei2,latticenuclei3}, 
just with Pionless EFT replaced by Chiral EFT.

After describing the lattice world in Sec. \ref{sec:0} and
reviewing Pionless EFT in Sec. \ref{sec:1},
the main results for lattice nuclei are summarized in  Sec. \ref{sec:2}.
An outlook is offered in Sec. \ref{sec:3}.
I hope Hussein is pleased with another unexpected
gift offered by weakly bound systems.

\section{EFT and QCD}
\label{sec:0}

It is intuitively clear, and supported by our experience in physics, 
that only certain degrees of freedom and symmetries
are relevant at a given distance scale.
Less obvious, but equally supported by the existence
of virtual processes in quantum mechanics, 
is that all interactions allowed by symmetry
take place among the relevant degrees of freedom. 
EFT is simply the framework that incorporates these facts.
To be predictive,
the basic assumption (``naturalness'') is that once a few scales are 
identified,
the infinite number of interaction strengths, 
or ``low-energy constants'' (LECs),
can be written as combinations
of these scales, times numbers of ${\cal O}(1)$.
As a consequence, observables, which can be expressed through
$T$ matrices for various processes, can be obtained as
controlled expansions in $Q/M$, where $Q$ represents the momenta of
interest and comparable mass scales, and $M$, high mass scales.
(I use natural units where $\hbar=c=1$.)
Because all interactions are included, observables are renormalization-group
(RG) invariant, that is, independent
of the arbitrary regularization procedure used to separate
explicit from implicit degrees of freedom.
For a basic introduction to EFT, see Ref. \cite{mylectures}.

Nuclear physics is well described in terms of nucleons subject
to (possibly approximate) Lorentz, (possibly approximate) baryon-number,
(approximate) time-reversal, and (approximate) parity invariance.
QCD for the two lightest quark flavors, as relevant
for nuclear physics, has essentially two separate scales: 
\begin{enumerate}
\item
$M_{QCD}\sim 1$ GeV,
where the coupling constant
of QCD formulated in terms of quarks and gluons 
becomes large.
$M_{QCD}$ sets the scale for hadronic masses, including the nucleon mass $m_N$,
and for $4\pi f_\pi$, where $f_\pi$ is the radius of the 
``chiral circle'' formed by the set of 
$SU(2)_L\times SU(2)_R\sim SO(4)$ minima of the QCD effective
potential. Picking one of these minima 
leads to spontaneous symmetry breaking, the emergence of 
pions as Goldstone bosons, and the manifestation
of $f_\pi\simeq 90$ MeV as the pion decay constant.
\item
$\bar m \sim 5$ MeV, the average quark mass, which explicitly breaks
chiral symmetry, creates an absolute minimum of the QCD effective
potential, and endows pions with a mass $m_\pi^2={\cal O}(M_{QCD}\bar m)$.
$\bar m$ also affects most other quantities, including $m_N$.
We can trade $\bar m$ for $m_\pi$.
\end{enumerate}

Contrary to real experiments, LQCD simulations can probe worlds
where $m_\pi$ takes different values. In fact, the high cost of
light quarks roaming the lattice constrains present calculations
to large values of $m_\pi$. While it certainly is a disadvantage
that LQCD cannot reach realistic values yet, one can turn 
a disadvantage into an advantage
by learning how nuclear physics depends on $m_\pi$.

Existing laboratory and LQCD data for nucleon masses
and light-nuclear binding energies are summarized in Tab. \ref{tbl:Data}.
I also list the EFT results described in Secs. \ref{sec:1} and \ref{sec:2}.
The LQCD calculations are performed with equal light quark masses
and no photons, hence isospin symmetry is exact and
the binding energy $B_A(S,I)$ is determined by the nucleon number $A$
and the spin-isospin combination $S,I$ of the state.
It did not have to be so, but masses and binding energies
in Tab. \ref{tbl:Data} increase more or less monotonically with $m_\pi$.
Note that Ref. \cite{lqcd_nn_force} finds no 
bound states
in a large range of pion masses that includes the values in 
Tab. \ref{tbl:Data}.
Since the model-independence of some
of its results, obtained through a non-observable
potential, remains a question mark, I do not consider
these data here. 

\begin{table}[bt]
\begin{center}
\caption{Neutron and proton masses, and 
binding energies of the lightest nuclei
at various values of the pion mass. All entries are in MeV.
The first column summarizes experimental data,
the third \cite{Yamazaki15}, fourth \cite{Yamazaki12} 
and fifth \cite{NPLQCD13a} columns give LQCD data,
and the second \cite{6N} and sixth \cite{latticenuclei} columns show EFT input 
(marked with $*$) and results.
The EFT calculations are discussed in Secs. \ref{sec:1} and \ref{sec:2}.
}
\label{tbl:Data}
\vspace{0.2cm}
\begin{small}
\begin{tabular}{c||cc|c|c|cc}
\hline
$m_{\pi}$ &  $140$ &  $140$  & $300$  & $510$ & $805$ & $805$ \\
Nucleus & [nature] & \cite{6N} & \cite{Yamazaki15} & \cite{Yamazaki12} 
& \cite{NPLQCD13a} & \cite{latticenuclei}\\
\hline
\hline
n     &  939.6 &  939.0 $*$ & 1053 &  1320     & 1634  & 1634 $*$ \\ 
p     &  938.3 &  939.0     & 1053 &  1320     & 1634  & 1634   \\
\hline                  
$^2$n & --- & --- & 8.5 $\pm$ 0.7 $^{+2.2}_{-0.4}$& 7.4 $\pm$ 1.4 
& 15.9 $\pm$ 3.8  & 15.9 $\pm$ 3.8 $*$ \\
$^2$H & 2.224 & 2.224 $*$ & 14.5 $\pm$ 0.7 $^{+2.4}_{-0.7}$& 11.5 $\pm$ 1.3 
& 19.5 $\pm$ 4.8 & 19.5 $\pm$ 4.8 $*$ \\
$^3$n & ---  &   &   &     &     &   $<$ 12.1     \\
$^3$H & 8.482 & 8.482 $*$ & 21.7 $\pm$ 1.2 $^{+5.7}_{-1.6}$& 20.3 $\pm$ 4.5 
& 53.9 $\pm$ 10.7 & 53.9 $\pm$ 10.7 $*$ \\
$^3$He& 7.718 &   8.482 & 21.7 $\pm$ 1.2 $^{+5.7}_{-1.6}$& 20.3 $\pm$ 4.5 
& 53.9 $\pm$ 10.7 & 53.9 $\pm$ 10.7\\
$^4$He& 28.30 & 28.30 $*$ & 47$\pm$ 7 $^{+11}_{-9}$& 43.0 $\pm$ 14.4 
& 107.0 $\pm$ 24.2 & 89 $\pm$ 36\\
$^4$He$^*$ & 8.09 & 10 $\pm$ 3 &   &    &    & $<$ 43.2  \\
$^5$He& 27.50 &      &     &   &    &  98 $\pm$ 39\\
$^5$Li& 26.61 &      &     &   &    &  98 $\pm$ 39\\
$^6$Li& 32.00 & 23 $\pm$ 7 &   &      &          & 122 $\pm$ 50\\
\hline
\end{tabular}
\end{small}
\end{center}
\end{table}

{}From the experimental
and LQCD data in Tab. \ref{tbl:Data}, we can infer
the 
relevant momentum scales for real and lattice nuclei,
which I list 
in Tab. \ref{tbl:MomScales}.
Besides pion and nucleon masses,
the momentum associated with the excitation of the lowest
baryon, the Delta isobar, in two-nucleon scattering \cite{savagedeltas}
is also given, using the LQCD data for the Delta mass $m_\Delta$
compiled in Ref. \cite{Alvarez13}.
As one can see, for the three values of the pion mass
the typical nuclear momenta $\sqrt{2m_N B_A/A}$ for 
$A=2,3,4$ are much smaller than $m_N$, suggesting that
a description in terms of non-relativistic nucleons 
is always appropriate.

\begin{table}[bt]
\begin{center} 
\caption{Momentum scales for various quark masses:
nucleon mass, effective momentum for Delta isobar excitation,
pion mass, and effective binding momenta for $S$-shell nuclei.
All entries are in MeV.} 
\label{tbl:MomScales}
\vspace{0.2cm}
\begin{small}
\begin{tabular}{c|ccc}
$m_N$                      & $1000$      & $1300$      & $1600$    \\
$\sqrt{2m_N(m_\Delta-m_N)}$  & $750 $      & $900$       &  $800$    \\
$m_{\pi}$                   & $140$       &  $500$      &  $800$    \\
$\sqrt{2m_N B_A/A}$ ($A$=2-4) & $50$-$110$ &  $130$-$170$ &  $190$-$300$  \\
\end{tabular}
\end{small}
\end{center}
\end{table}

EFTs for non-relativistic nucleons have some simple, useful features.
Pair creation is a short-range effect
and the theory can be formulated in terms of Pauli spinors 
representing forward propagation in time. 
The EFT Lagrangian contains the usual
non-relativistic kinetic terms in lowest order, with relativistic corrections
implemented at higher orders in a $Q/m_N$ expansion.
Nucleon energies are of ${\cal O}(Q^2/m_N)$.
Loops with antinucleons never need to be considered explicitly,
and an $N$-body force does not affect the 
$n$-body system for $n<N$.
At the $N$-nucleon level, 
all possible operators involving up to $2N$ nucleon fields
are included, with an increasing number of derivatives,
These interactions are highly singular and require regularization
via some ultraviolet (UV) cutoff $\Lambda$.
The explicit dependence on non-negative powers of $\Lambda$
coming from loops is eliminated by renormalization of the LECs.
Once (and only once) this is done at a given order,
an integration over momenta in intermediate states 
contributes a factor ${\cal O}(Q^3/4\pi)$ to the $T$ matrix.
These factors of $Q$, $m_N$ and $4\pi$, together with 
the sizes of the renormalized LECs, are the ingredients
to build the $Q/M$ expansion.

In nature, there is a large separation of scales, $m_\pi \ll M_{QCD}$.
A nucleon thus consists of an outer cloud of pions at
distances $\sim 1/m_\pi$ surrounding
an unresolved, dense core of
size $\sim 1/M_{QCD}$. 
Empirically, nuclear sizes scale as 
$R_A\sim A^{1/3}r_0$, where $r_0\sim 1.2$ fm is possibly set by a combination
of $f_\pi$ and  $m_\pi$. 
Nuclei are expected to be large on the $1/M_{QCD}$ distance scale because
of the two effects:
the size of the pion cloud around each nucleon and the piling
up of nucleons. 
LQCD has to fight both effects to contain nuclei
within the lattice length $L\simge R_A$, while
striving for a much smaller short-distance regulator
in the form of a lattice spacing $b\simle 1/M_{QCD}$. 
EFT offers a strategy to extrapolate QCD to the large
distances involved in nuclear physics: 
{\it i)} calculate with LQCD $A$-nucleon observables for $A=2,3,4$;
{\it ii)} calculate the same observables with EFT and match LQCD, thus 
determining the LECs;
and {\it iii)} solve the EFT for $A\ge 5$
using the powerful ``{\it ab initio}'' methods that have been
developed in recent years, such as
the no-core shell model (NCSM) \cite{NCSM},
the effective-interaction hyperspherical harmonics (EIHH) \cite{EIHH},
and 
the auxiliary-field diffusion Monte Carlo (AFDMC) \cite{AFDMC1}
methods.

For momenta $Q\sim m_\pi\ll M_{QCD}$, one can formulate an EFT,
Chiral EFT \cite{paulo,nonEFT},  which includes,
in addition to nucleons, also pions and the lowest nucleon excitations.
In this EFT, $M\sim M_{QCD}$, and 
one treats the inner nucleon cloud in a multipole-type
expansion. 
The approximate chiral symmetry of QCD plays a crucial role,
because it ensures that pions couple weakly at low momenta, 
which gives rise to a loop,
or equivalently density, expansion for the larger, more sparse pion cloud.
Chiral EFT allows one to, in principle, calculate the dependence of low-energy 
nuclear observables on $m_\pi$ \cite{BBSvK}. Unfortunately, however, 
an RG-invariant formulation of Chiral EFT is still work in progress 
\cite{RGIChiEFT}.
This is not too serious a problem in the sense that 
it is doubtful that the Chiral EFT expansion holds at pion masses
explored so far by LQCD. 
For example, studies of the convergence of Chiral EFT  
for $A=0$ suggest a breakdown of the expansion
at a pion mass no larger than 500 MeV \cite{Durr:2014oba}.
When smaller pion masses can be reached
and a proper formulation well developed,
Chiral EFT could be used as a tool for extrapolation 
of nuclear quantities in $m_\pi$, as it is already for meson and one-nucleon
observables.
Chiral EFT LECs would then be determined from LQCD instead
of experimental data, and a solution of Chiral EFT used
to extrapolate LQCD to larger $A$.

\section{Pionless EFT}
\label{sec:1}

In fact, Tab. \ref{tbl:MomScales} suggests that 
the extrapolation to larger $A$ can already be performed.
For all available pion masses, the typical nuclear momentum 
is not only much smaller than $m_N$, it is also smaller than
$m_\pi$. This is already true in nature
and, as the pion mass increases, pion effects become more short-ranged
relative to nuclear distances.
Assuming that the Delta continues to be the lowest baryon excitation,
any effects from other hadrons are also short-ranged.

Thus, at momenta $Q\ll m_\pi$, nucleons should suffice as explicit
degrees of freedom.
In the appropriate EFT, Pionless EFT \cite{paulo}, 
$M\sim m_\pi$ and {\it all} interactions are of contact type,
which in coordinate space give (renormalized
versions of) delta functions and derivatives.
In nature, Pionless EFT describes well the properties of
low-energy scattering and bound states
for $A=2$ \cite{2N}, 
$3$ \cite{3N,3Nscatt}, 
$4$ \cite{4N,4Nscatt}, 
and even (but less well) $6$ \cite{6N}.
One expects Pionless EFT to breakdown at some point
as nuclei get denser, but its reach is presently unknown.  

In the two-nucleon sector,
there are two independent non-derivative contact interactions, with 
renormalized LECs 
$C_{01}={\cal O}(4\pi a_2(0,1)/m_N)$ and 
$C_{10}={\cal O}(4\pi a_2(1,0)/m_N)$, in terms of
the $^1S_0$ and $^3S_1$ scattering lengths, respectively
$a_2(0,1)$ and $a_2(1,0)$.
Together with the general estimates for nucleon energies
and loop integrals given above,
each iteration of this potential in the $T$ matrix
yields a factor ${\cal O}(Q|a_2|)$.
When $|a_2|\gg 1/M$ as a consequence of a shallow pole
in the $T$ matrix, one needs to include
all iterations for $Q\simge 1/|a_2|$, 
namely solve the corresponding Schr\"odinger equation exactly.
It is easy to show that the dangerous UV regulator dependence
can be eliminated
if the bare LECs $C_{SI}(\Lambda)\propto 1/\Lambda$.
In coordinate space one can understand this by noticing
that a delta function is a $\Lambda^3/4\pi$ singularity and
overwhelms the kinetic term, which grows at most as $\Lambda^2/m_N$,
unless the associated LEC goes as $4\pi/(m_N\Lambda)$.
The relative $(a_2\Lambda)^{-1}$ corrections in $C_{SI}(\Lambda)$ provide then
just the balance necessary for a low-energy (real or virtual) bound state
with binding momentum ${\cal O}(1/a_2)$.
The regularization procedure leaves 
behind a relative error of ${\cal O}(Q/\Lambda)$,
which can be made arbitrarily small by taking $\Lambda$
arbitrarily large. In leading order (LO) we obtain the first term in the 
effective-range expansion (ERE) of the $T$ matrix.
The residual cutoff dependence can be removed 
by the two-derivative interactions present in the same channels,
which, as a consequence, 
have a natural relative size ${\cal O}(Q/M)$ with respect to LO.
That means they make up the next-to-leading order (NLO), 
their renormalized LECs 
being of ${\cal O}(4\pi a_2^2r_2/m_N)$, with
effective ranges $|r_2|={\cal O}(1/M)$.
The argument can be generalized to more-derivative terms in the $S$ waves,
which contribute at progressively higher orders.
In higher waves, where there seem to be no shallow poles,
all LECs scale with $M$ according to their canonical dimension,
which means that no other wave needs to be considered up to N$^2$LO.
RG invariance requires that, as allowed by their small
relative sizes, subleading orders be treated in perturbation theory.
The gory details are spelled out in Ref. \cite{piless2}.

In this way, Pionless EFT generates an expansion of the two-nucleon 
amplitude equivalent \cite{piless2} 
to the ERE.
At LO, only the two non-derivative contact interactions
need to be included.
Each LEC is determined by one datum,
say the scattering length or the binding momentum of the shallow pole
(which differ only by higher-order terms).
At NLO the two $S$-wave two-derivative contact interactions
need to be included in first distorted-wave Born approximation,
and each new LEC requires another datum, say the effective range. 
A nice description of two-nucleon scattering 
at physical quark mass was found in Ref. \cite{2N}.
In the regime $1/|a_2|\ll Q \ll M$, the amplitude is approximately
scale invariant and $SU(4)$ spin-isospin symmetric \cite{Mehen:1999qs}.

Nuclear-physics folklore would suggest that few-nucleon forces are
of higher order. Indeed, in certain channels such as
$^4S_{3/2}$ neutron-deuteron scattering, high accuracy can be obtained in the
first orders \cite{3Nscatt}. However, in the $^2S_{1/2}$ 
channel \cite{3N}  
(and for three
bosons in relative $S$ waves \cite{piless3}), the only way to eliminate 
non-negative powers of $\Lambda$ 
(in particular, the ``Thomas collapse'' \cite{thomas}
of the ground state)
is to ensure that the non-derivative
six-nucleon operator (there is only one) is present in LO.
Again the cutoff dependence of its bare LEC,
$D(\Lambda)\propto 1/\Lambda^4$,
can be understood
by a simple coordinate-space argument. When going from two to three bodies,
the delicate balance between kinetic terms and two-body
contact is destroyed because the number of kinetic terms doubles while
the number of pair-wise interactions triples. The result is the Thomas 
collapse where the two-body attraction wins and leads to a three-body
binding energy that grows as $\Lambda^2/m_N$. 
This growth has to be canceled by the contact three-body
interaction; since it involves two delta functions, or
$\Lambda^6/(4\pi)^2$, the bare LEC should be roughly
$(4\pi)^2/(m_N\Lambda^4)$.
The cutoff dependence of $D$ is still more complicated, though.
At very low cutoffs the three-body force might be attractive or repulsive,
but once $\Lambda^2/m_N$ exceeds the binding energy of the ground state,
the three-body force must be increasingly repulsive to prevent the collapse.
With a regulator procedure that preserves the approximate scale invariance
of the two-body subsystems,
at a critical $\Lambda$
one can maintain the binding energy fixed only
at the cost of making the three-body
force attractive and accreting a very deep bound state. As the cutoff
increases past critical, the two-body attraction continues to increase
and the three-body force gets less attractive, till it becomes
repulsive and a new cycle begins. In fact \cite{piless3,3N}
the bare LEC is on an RG limit cycle,
approximate scale invariance is reduced
to an approximate {\it discrete} scale invariance,
and there is a tower of approximately geometric three-body ``Efimov 
states'' \cite{efimov}. 
Varying the renormalized $D={\cal O}((4\pi)^2 a_2^4/m_N)$
shifts the position of the tower and changes
the three-body scattering length, leading to a correlation
known as the Phillips line \cite{phillips}. 
In nuclear physics the explicit breaking
of scale invariance is such that the ``tower'' consists of a single
state, the triton.
The two-derivative three-body force first
appears at N$^2$LO \cite{piless3,3N}.

Calculations \cite{piless4,4N}, which are however somewhat limited in cutoff 
variation,
indicate that four-body observables do not display non-negative powers 
of $\Lambda$ up to NLO, in the absence of four-body forces. 
This is perhaps not surprising
since the three-body force is effectively repulsive
and the number of triplets grows faster than doublets.
With fixed two-nucleon input, 
variation in $D$ leads to a correlation between 
four- and three-body binding energies, the 
``Tjon line'' \cite{tjon}, which passes
close to the experimental point.
Pionless EFT thus successfully postdicts the alpha-particle 
binding energy \cite{4N},
and scattering can be calculated as well \cite{4Nscatt}.
Notice, however, that 
the arguments above leave open the possibility of cutoff dependence
in the regions where the three-body force is attractive, and indeed
Ref. \cite{tobias} found
sensitivity in four-body properties to a four-body scale. 
As far as I can see, there is yet no compelling argument that
this sensitivity is due to an {\it LO (or even NLO)} force. 
I will assume
that the dominant four-body force, presumably coming
from the single non-derivative eight-nucleon operator, 
first contributes beyond NLO.
Since, thanks to the Pauli principle,
five- or more-body forces involve at least two derivatives,
they are likely of even higher order.

It is probably safe to assume
that, up to NLO, Pionless EFT is renormalizable 
with, besides two-body forces,
a single non-derivative three-body force.
Its LEC is determined by one three-body datum,
say the neutron-deuteron $^2S_{1/2}$ scattering length
or the triton binding energy.
Thus at LO (NLO) three (five) data are needed as input,
in addition to the nucleon mass, and everything else is a prediction.
Pionless EFT is not {\it just} the ERE; it is the extension
to few-body systems that preserves model independence.
Pionless EFT accounts for a series of apparently
unrelated, qualitatively unique phenomena,
such as the Thomas collapse, Efimov states, the Phillips
and Tjon lines, and presumably similar correlations for
bigger systems.
With a few adaptations, it applies to other systems characterized
by a small ratio $r_2/a_2$ ---see, {\it e.g.} Refs. \cite{6He,3atoms,3fesh}.
The approximate discrete scale invariance has striking consequences for
the spectrum of few-boson systems \cite{mario},
where a state in Efimov's three-body tower generates pairs of 
``image'' states in bigger systems. 
The alpha-particle
ground and excited states can likely be interpreted this way. 
Moreover, since the LO three-body force is $SU(4)$ symmetric, 
Pionless EFT provides a justification \cite{piless3,3N}
for the approximate $SU(4)$ symmetry proposed by Wigner \cite{wigner}.

It is unfortunately still unclear how far up the nuclear chart 
this EFT can be pushed. The only calculation \cite{6N}
beyond the four-nucleon system
was based on the NCSM \cite{NCSM},
when the EFT Hamiltonian was diagonalized in a harmonic-oscillator basis.
This basis has a natural 
UV cutoff in the form of a maximum
allowed number of shells.
It contains also an infrared (IR) cutoff provided by
the spacing between shells.
In the simplest approach, the LECs are fitted to the 
experimental binding energies of the lightest nuclei for every cutoff pair,
and 
binding energies for larger nuclei are calculated 
and extrapolated to large UV and small IR cutoffs.
Results of an LO calculation \cite{6N}, where the deuteron, triton, and
alpha-particle ground-state energies were used as input
in addition to the nucleon mass,
are shown in Tab. \ref{tbl:Data}.
The input data are indicated by a ``$*$'' in Tab. \ref{tbl:Data}.
Estimating the error as 30\% from $r_2/a_2$ in the $^3S_1$ channel,
one sees that the excited state of the alpha particle is postdicted
very well, while $^6$Li is barely consistent.
However, the error could be as large as 80\% if we consider
the ratio in Tab. \ref{tbl:MomScales}
between alpha-particle momentum and pion mass.
Higher-order calculations are clearly needed.

\section{EFT for Lattice Nuclei}
\label{sec:2}

As noted in Ref. \cite{latticenuclei}, 
the widening gap shown in Tab. \ref{tbl:MomScales}
between pion mass and typical nuclear momentum
implies Pionless EFT should work better at larger pion masses.
In the first calculation ever to fit lattice {\it nuclear} data, 
Ref. \cite{latticenuclei} used the nucleon mass and light-nuclear
binding energies at the highest pion-mass value,
$m_{\pi}=805$ MeV, from the NPLQCD collaboration \cite{NPLQCD13a}
as input for Pionless EFT in LO.
A calculation using the $m_{\pi}=510$ MeV
binding energies from Ref. \cite{Yamazaki12} is in 
progress \cite{latticenuclei2}.
The existence of a dineutron bound state allows 
the use of its binding energy instead of the alpha-particle's
as input.

{}From Tab. \ref{tbl:MomScales} a conservative
estimate for the $Q/m_\pi$ expansion parameter at $m_{\pi}=805$ MeV 
is 40\%. 
Two-nucleon scattering lengths and effective ranges are also
available at this pion mass
\cite{NPLQCD13b},
and are consistent with an
almost degenerate double bound-state pole in the $T$ matrix
of each $S$ wave, 
which is thought to be 
incompatible with a short-range non-relativistic potential
\cite{russians}.
References \cite{latticenuclei,latticenuclei2} assume 
$|r_2|={\cal O}(1/m_\pi)$, as for physical quark masses.
If one uses the ratio $r_2/a_2$, the error estimate for LO is instead 50\%.
A test of convergence will have to await
an NLO calculation.

In a renormalizable theory, only convenience guides the choice of regulator.
In the present case, we want, as in Ref. \cite{Gezerlis13},  
a local potential that allows the use of many-body techniques
that cannot handle non-local interactions well, such as 
AFDMC \cite{AFDMC1}. 
This can be achieved with a regulator function $f(\vec{q}^2/\Lambda^2)$
in the momentum transfer $\vec{q}$,
or its Fourier transform $F(r^2\Lambda^2)$ in terms of the 
radial coordinate $r$.
Ref. \cite{latticenuclei} employed
two forms, $f_{n}(x)= \exp(-x^{2n})$ with $n=1,2$,
which get increasingly closer to a sharp regulator.
The isospin-symmetric Hamiltonian can be written in coordinate space as
\begin{eqnarray}
 H &=& - \frac{1}{2 m_N}\sum_i \nabla^2_i
 +\frac{1}{4}\sum_{i<j} 
  \left[3C_{10}(\Lambda)+C_{01}(\Lambda)
  +\left(C_{10}(\Lambda)-C_{01}(\Lambda)\right)\vec{\sigma}_i\cdot\vec{\sigma}_j 
  \right] F(r_{ij}^2\Lambda^2) 
\nonumber\\
&& +\sum_{i<j<k}\sum_{cyc} D(\Lambda) \, \boldtau_i\cdot\boldtau_j\,
  F(r_{ik}^2\Lambda^2)  F(r_{jk}^2\Lambda^2)  
+\ldots, 
\label{Heff}
\end{eqnarray}
where $\vec{\sigma}_i/2$ ($\boldtau_i/2$) is the spin (isospin)
of nucleon $i$,
$\sum_{cyc}$ stands for the cyclic permutation of a particle triplet $(ijk)$,
and ``$\ldots"$ for terms containing 
more derivatives and/or more-body forces.
The LECs $m_N$, $C_{10}(\Lambda)$, $C_{01}(\Lambda)$, 
$D(\Lambda)$, {\it etc.}
depend on $m_{\pi}$, since pions are part of the short-distance physics
not included explicitly.
The $m_\pi$ dependence of two- and three-nucleon observables
in Pionless EFT has been studied with input from Chiral EFT
in Ref. \cite{hammeretal}.

The two-nucleon Schr\"odinger equation was solved \cite{latticenuclei}
for the LO Hamiltonian with the Numerov method,
and $C_{10}(\Lambda)$ and $C_{01}(\Lambda)$ fitted to the deuteron $B_2(1,0)$ and
dineutron $B_2(0,1)$ binding energies \cite{NPLQCD13a},
respectively.
The cutoff dependence of the LECs is found to be
qualitatively similar to other regulators \cite{piless2,2N}:
$C_{SI}(\Lambda)\Lambda$ approaches a regulator-specific constant 
at a rate determined by $\sqrt{m_N B_2(S,I)}$.
For large cutoffs one should have in LO
$a_{2}(1,0)\approx 1/\sqrt{m_N B_2(1,0)}\simeq 1.12$ fm.
For cutoff variation in the range $2$-$14$ fm$^{-1}$,
Ref. \cite{latticenuclei} finds explicitly
$a_{2}(1,0)=(1.2\pm 0.5)$ fm
($(1.1\pm 0.1)$ fm) with the regulator $f_1$ ($f_2$).
For comparison, LQCD gives $a_{2}(1,0)=1.82^{+0.14+0.17}_{-0.13-0.12}$ fm
directly \cite{NPLQCD13b}.
The situation is similar in the $^1S_0$ channel.

For systems with $3\le A\le 6$ nucleons the 
Schr\"odinger equation was solved \cite{latticenuclei} with
the EIHH method, where
the wavefunction is expanded into a set of
antisymmetrized hyperspherical-harmonic spin-isospin states.
Convergence is controlled by the hyper-angular quantum number
$K_{max}$, results being obtained by extrapolation to the limit
$K_{max}\to \infty$ \cite{EIHH}.
The corresponding error was estimated 
to be smaller (for the lighter systems, much smaller) than
the EFT truncation error.
For systems with $A\ge 4$ the AFDMC method was also used.
In this technique \cite{AFDMC1},
the ground-state energies are projected 
from an arbitrary initial state by means of a stochastic imaginary-time 
propagation. The numerical simulations are simplified
by the introduction of auxiliary fields 
via a Hubbard-Stratonovich transformation.  
In all these calculations \cite{latticenuclei}, the regulator employed  
was $f_1$ with $2\leq \Lambda \; \rm{fm} \leq 8$. 

The LEC $D(\Lambda)$ was determined \cite{latticenuclei}
imposing that the $^3$H/$^3$He binding energy $B_3$ is reproduced
at any value of $\Lambda$.
It was found that $D(\Lambda)\Lambda^4$ approaches a finite
limit, as for other regulators \cite{piless3,3N}.
The limit-cycle behavior is not seen, 
as in other cases when the number of three-body bound states
is kept fixed with regulators that do not preserve
the approximate scale invariance of the two-body subsystems,
{\it e.g.} Ref. \cite{Rotureau:2011vf}.
With LECs thus fixed, a complete LO potential is available 
to predict other properties of lattice nuclei.

The four-nucleon system, solved  \cite{latticenuclei}
with both EIHH and AFDMC methods,
provides a consistency check between the two {\it ab initio} methods, and 
between them and LQCD.
The two {\it ab initio} methods produced results for the 
$^4$He binding energy $B_4$
that agree well within the (large) LQCD error.
In either case, $B_4$ was found to depend only weakly on
the cutoff, changing by about
20\% when $\Lambda$ grows by a factor of 4. 
Over a wide cutoff range
the EFT prediction reproduces the
LQCD result within its error, evidence that
the EFT in LO captures the essence of the strong-interaction dynamics.
As $D$ varies (at fixed $\Lambda$) within the error bars
of $B_3$, $B_4$ also changes within its error bars.
The estimate of a 40\% error in LO EFT is likely conservative, indeed.

The power of EFT is the relative ease with which it can
be solved for more-body systems.
Binding energies for $A\ge 5$ are predictions 
that extend LQCD into new territory.
Using $\Lambda=2\,\rm{fm}^{-1}$,
the authors of Ref. \cite{latticenuclei} searched unsuccessfully 
for excited states in $A=2,3,4$ systems. 
Similarly, they found no evidence of $^3$n droplets,
for which the
ground-state binding energy coincided with the two-body threshold.
Results \cite{latticenuclei}
for the $A=5,6$ ground states at $\Lambda=2\,\rm{fm}^{-1}$
are shown in Tab. \ref{tbl:Data},
with errors estimated from the EFT truncation.
For $^5$He a bound state 
with binding energy $B_5=98.2$ MeV for $\Lambda=2\;\rm{fm}^{-1}$ 
coincided with the four-body threshold
for $\Lambda=4\,\rm{fm}^{-1}$.
The $^6$Li ground state
for $\Lambda=2\, \rm{fm}^{-1}$ was found at $B_6\approx 122\,\rm{MeV}$.
In this case the error in $K_{max}$ extrapolation 
was about 3 MeV, which
is somewhat larger than for lighter systems but still
small compared with input and truncation errors.
Calculations with AFDMC at larger $A$ are in progress \cite{latticenuclei3}.

These results are afflicted by considerable error bars.
Even though the 40\% assigned to the EFT expansion is likely
an overestimate, the LQCD input itself has large uncertainties
of about 25\%.
With this caveat, the trend of the results is surprising. 
There is a qualitative difference
with $A=2$ at the physical pion mass, because 
the dineutron is bound at larger masses.
(For the effects of the dineutron scattering length
on light nuclei in Pionless EFT at physical pion mass, see
Ref. \cite{nnscattlength}.)
But this is consistent with other binding energies,
which are all larger, and
all larger by roughly similar amounts.
The gap at $A=5$, familiar in
nature, seems to survive the increase in pion mass.
And $B_6/6 \approx 20$ MeV, similar to 
lattice $^4$He for which $B_4/4 \approx 25$ MeV.
This suggests that nuclear saturation 
might not be tremendously sensitive to the pion mass.
Overall, it seems that the lattice world at $m_{\pi}=805$ MeV
is not that different from our own, 
with $B_A/A$ scaled by a factor 4 or 5. 

\section{Outlook}
\label{sec:3}

Natural light ($A\le 4$) nuclei are halo-type systems in the sense that
they have sizes large compared the range of the force,
and are thus described by Pionless EFT. 
One might expect this feature to result from fine-tuning,
and to find very different worlds at unphysical pion masses.
Surprisingly, the first LQCD calculations and their EFT extrapolations
seem to suggest the opposite.
Perhaps pions do not play as decisive a role in low-energy QCD 
as we are used to think, and 
some of the defining properties of nuclei 
are relatively insensitive to the value of the pion mass.
If this is true, it should have implications for the
use of nuclei in tests of the variability of fundamental constants
\cite{funcons}.

Of course, at this point these are only hints.
The same exercise can be, and is being \cite{latticenuclei2}, carried out 
in LO with the LQCD data at $m_\pi=500$ MeV \cite{Yamazaki12}. 
EFT extrapolations can be repeated at other pion masses 
as LQCD results appear.
More urgently, EFT calculations 
need to be performed for $A\ge 4$ at higher
cutoffs to confirm renormalizability, even at the physical pion mass.
The scattering lengths and effective ranges at $m_\pi=800$ MeV \cite{NPLQCD13b}
give just enough input for an NLO analysis, which should
allow stronger statements about the convergence of Pionless EFT
at unphysical pion masses. 
Finally, at all values of pion mass, one should increase $A$ 
\cite{latticenuclei3}
to confirm trends in $B_A/A$ and to seek the limit
of applicability of the EFT.

On a longer time frame, as pion masses in LQCD drop sufficiently,
one can use Chiral EFT to extrapolate further down in pion mass
and fully predict real nuclei,
following the same steps as in Refs. 
\cite{latticenuclei,latticenuclei2,latticenuclei3}.
It is a lot to do, but it promises to fulfill
a longstanding dream of nuclear effective field theorists,
teaching us much about
the connection between QCD and nuclear physics.

\section*{Acknowledgments}
I would like to thank my collaborators Nir Barnea, Lorenzo Contessi,
Doron Gazit, Johannes Kirscher, and Francesco Pederiva for a fun, on-going
trip into an unfamiliar world.
Thanks, in particular, to Johannes for insightful comments on the manuscript.
I am also grateful to Tobias Frederico, Mahir Hussein, and Lauro Tomio
for the invitation to a great meeting.
This material is based upon work supported 
in part by the U.S. Department of Energy, 
Office of Science, Office of Nuclear Physics, 
under Award Number DE-FG02-04ER41338.


\begin{thebibliography}{3}

\bibitem{Canto}
L.F. Canto, P.R.S. Gomes, R. Donangelo, and M.S. Hussein, 
Phys. Rept. {\bf 424} (2006) 1.

\bibitem{Bertulani:1993qt}
C.A. Bertulani, L.F. Canto, and M.S.~Hussein,
Phys. Rept. {\bf 226} (1993) 281.

\bibitem{Timmermans}
E. Timmermans, P. Tommasini, M.S. Hussein, and A.K. Kerman,
Phys. Rept. {\bf 315} (1999) 199.

\bibitem{latticenuclei}
N. Barnea, L. Contessi, D. Gazit, F. Pederiva, and U. van Kolck,
Phys. Rev. Lett. {\bf 114} (2015) 052501.

\bibitem{latticenuclei2}
J. Kirscher {\it et al.}, in preparation.

\bibitem{latticenuclei3}
L. Contessi {\it et al.}, in progress.

\bibitem{Yamazaki12} 
T. Yamazaki, K.-i. Ishikawa, Y. Kuramashi, and A. Ukawa, 
Phys. Rev. D {\bf 86} (2012) 074514.

\bibitem{NPLQCD13a} 
S.R. Beane {\it et  al.} (NPLQCD Collaboration), 
Phys. Rev. D {\bf 87} (2013) 034506.

\bibitem{NPLQCD13b} 
S.R. Beane {\it et  al.} (NPLQCD Collaboration), 
Phys. Rev. C {\bf 88} (2013) 024003. 

\bibitem{Yamazaki15}
T. Yamazaki, K.-i. Ishikawa, Y. Kuramashi and A. Ukawa,
arXiv:1502.04182 [hep-lat].

\bibitem{piless2} 
U. van Kolck, 
Nucl. Phys. A {\bf 645} (1999) 273.

\bibitem{piless3} 
P.F. Bedaque, H.-W. Hammer, and U. van Kolck,
Nucl. Phys. A {\bf 646} (1999) 444;
C. Ji, D.R. Phillips, and L. Platter,
Annals Phys. {\bf 327} (2012) 1803.

\bibitem{piless4} 
L. Platter, H.-W. Hammer, and U.-G. Mei{\ss}ner,
Phys. Rev. A {\bf 70} (2004) 052101;
H.-W.~Hammer and L.~Platter,
Eur. Phys. J. A {\bf 32} (2007) 113.

\bibitem{paulo}
P.F. Bedaque and U. van Kolck,
Ann. Rev. Nucl. Part. Sci. {\bf 52} (2002) 339.

\bibitem{6He}
J. Rotureau and U. van Kolck,
Few-Body Syst. {\bf 54} (2013) 725;
C. Ji, C. Elster, and D.R. Phillips,
Phys. Rev. C {\bf 90} (2014) 044004.

\bibitem{3atoms}
P.F. Bedaque, H.-W. Hammer, and U. van Kolck,
Phys. Rev. Lett. {\bf 82} (1999) 463;
C. Ji and D.R. Phillips,
Few Body Syst. {\bf 54} (2013) 2317.

\bibitem{3fesh}
P.F. Bedaque, E. Braaten, and H.-W. Hammer,
Phys. Rev. Lett.  {\bf 85} (2000) 908;
C. Ji, D. Phillips, and L. Platter,
Europhys. Lett. {\bf 92} (2010) 13003.

\bibitem{2N}
J.-W. Chen, G. Rupak, and M.J. Savage,
Nucl. Phys. A {\bf 653} (1999) 386;
X. Kong and F. Ravndal,
Phys. Lett. B {\bf 450} (1999) 320.

\bibitem{3N}
P.F. Bedaque, H.-W. Hammer, and U. van Kolck,
Nucl. Phys. A {\bf 676} (2000) 357;
H.W. Grie{\ss}hammer,
Nucl. Phys. A {\bf 760} (2005) 110;
S.i. Ando and M.C. Birse,
J. Phys. G {\bf 37} (2010) 105108;
J. Vanasse, D.A. Egolf, J. Kerin, S. K\"onig, and R.P. Springer,
Phys. Rev. C {\bf 89} (2014) 064003.

\bibitem{3Nscatt}
P.F. Bedaque and U. van Kolck,
Phys. Lett. B {\bf 428} (1998) 221;
P.F. Bedaque, H.-W. Hammer, and U. van Kolck,
Phys. Rev. C {\bf 58} (1998) 641;
P.F. Bedaque and H.W. Grie{\ss}hammer,
Nucl. Phys. A {\bf 671} (2000) 357;
J. Vanasse,
Phys. Rev. C {\bf 88} (2013) 044001.

\bibitem{4N}
L. Platter, H.-W. Hammer, and U.-G. Mei{\ss}ner,
Phys. Lett. B {\bf 607} (2005) 254;
J. Kirscher, H.W. Grie{\ss}hammer, D. Shukla, and H.M. Hofmann,
Eur. Phys. J. A {\bf 44} (2010) 239.

\bibitem{4Nscatt}
J. Kirscher,
Phys. Lett. B {\bf 721} (2013) 335.

\bibitem{6N}
I. Stetcu, B.R. Barrett, and U. van Kolck,
Phys. Lett. B {\bf 653} (2007) 358.

\bibitem{nonEFT}
R. Machleidt and D.R. Entem,
Phys. Rept. {\bf 503} (2011) 1;
E. Epelbaum and U.-G. Mei{\ss}ner,
Ann. Rev. Nucl. Part. Sci. {\bf 62} (2012) 159.

\bibitem{BBSvK}
S.R. Beane, P.F. Bedaque, M.J. Savage, and U. van Kolck,
Nucl. Phys. A {\bf 700} (2002) 377;
S.R. Beane and M.J. Savage,
Nucl. Phys. A {\bf 717} (2003) 91.

\bibitem{mylectures}
U. van Kolck,
{\it Lect. Notes Phys.} {\bf 879} (2014) 123.

\bibitem{lqcd_nn_force}
T. Inoue {\it et  al.} (HAL QCD Collaboration),
Nucl. Phys. A {\bf 881} (2012) 28.

\bibitem{savagedeltas}
M.J. Savage,
Phys. Rev. C {\bf 55} (1997) 2185. 

\bibitem{Alvarez13} 
L. Alvarez-Ruso, T. Ledwig, J. Martin Camalich, and M.J. Vicente-Vacas,
Phys. Rev. D {\bf 88} (2013) 054507;
and references therein.

\bibitem{NCSM}
P. Navratil, J.P. Vary, and B.R. Barrett,
Phys. Rev. Lett. {\bf 84} (2000) 5728;
Phys. Rev. C {\bf 62} (2000) 054311.

\bibitem{EIHH} 
N. Barnea, W. Leidemann, and G. Orlandini,
Phys. Rev. C {\bf 61} (2000) 054001; 
Nucl. Phys. A {\bf 693} (2001) 565. 

\bibitem{AFDMC1}
K.E. Schmidt and S. Fantoni, 
Phys. Lett. B {\bf 446} (1999) 99;
S. Gandolfi {\it et al.}, 
Phys. Rev. C {\bf 79} (2009) 054005. 

\bibitem{RGIChiEFT}
A. Nogga, R.G.E. Timmermans, and U. van Kolck,
Phys. Rev. C {\bf 72} (2005) 054006;
M. Pav\'on Valderrama,
Phys. Rev. C {\bf 83} (2011) 024003;
Phys. Rev. C {\bf 84} (2011) 064002;
B. Long and C.J. Yang,
Phys. Rev. C {\bf 85} (2012) 034002;
Phys. Rev. C {\bf 86} (2012) 024001.

\bibitem{Durr:2014oba}
S. D\"urr,
arXiv:1412.6434 [hep-lat].

\bibitem{Mehen:1999qs}
T. Mehen, I.W. Stewart, and M.B. Wise,
Phys. Rev. Lett. {\bf 83} (1999) 931.

\bibitem{thomas}
L.H. Thomas,
Phys. Rev. {\bf 47} (1935) 903.

\bibitem{efimov}
V.N. Efimov,
Sov. J. Nucl. Phys. {\bf 12} (1971) 589;
Phys. Rev. C {\bf 47}, 1876 (1993).

\bibitem{phillips}
A.C. Phillips,
Nucl. Phys. A {\bf 107} (1968) 209.

\bibitem{tjon}
J.A. Tjon, 
Phys. Lett. B {\bf 56} (1975) 217.

\bibitem{tobias}
M.R. Hadizadeh, M.T. Yamashita, L. Tomio, A. Delfino, and T. Frederico,
Phys. Rev. Lett. {\bf 107} (2011) 135304;
T. Frederico, A. Delfino, M.R. Hadizadeh, L. Tomio, and M.T. Yamashita,
Few-Body Syst. {\bf 54} (2013) 559.

\bibitem{mario}
M. Gattobigio, A. Kievsky, and M. Viviani,
Phys. Rev. A {\bf 86} (2012) 042513;
A. Kievsky, N.K. Timofeyuk, and M. Gattobigio,
Phys. Rev. A {\bf 90} (2014) 032504.

\bibitem{wigner}
E. Wigner,
Phys. Rev. {\bf 51} (1937) 106.

\bibitem{russians}
Y.N. Demkov and G.F. Drukarev, 
Sov. Phys. JETP {\bf 22} (1966) 479.

\bibitem{Gezerlis13} 
A. Gezerlis {\it et al.}, 
Phys. Rev. Lett. {\bf 111} (2013) 032501.

\bibitem{hammeretal}
E. Braaten and H.-W. Hammer, 
Phys. Rev. Lett. {\bf 91} (2003) 102002;
H.-W. Hammer, D.R. Phillips, and L. Platter,
Eur. Phys. J. A {\bf 32} (2007) 335.

\bibitem{Rotureau:2011vf}
J. Rotureau, I. Stetcu, B.R. Barrett, and U. van Kolck,
Phys. Rev. C {\bf 85} (2012) 034003.

\bibitem{nnscattlength}
J. Kirscher and D.R. Phillips,
Phys. Rev. C {\bf 84} (2011) 054004;
H.-W. Hammer and S. K\"onig,
Phys. Lett. B {\bf 736} (2014) 208.

\bibitem{funcons}
P.F. Bedaque, T. Luu, and L. Platter,
Phys. Rev. C {\bf 83} (2011) 045803;
A. Coc, P. Descouvemont, K.A. Olive, J.-P. Uzan, and E. Vangioni,
Phys. Rev. D {\bf 86} (2012) 043529.

\end{thebibliography}
\end{document}